# Effect of Attachment Surface on Biofilm Response and the Dissipation of PAHs and nitrogen in Submerged Plant System under the application of Bactericide and Algaecide

Xiabinchen Zhao[a,b*], Liling Xia[c], Zhenhua Zhao[d]

[a] Jinling High School. Nanjing 210005, P.R. China

[b] Okemos High School, East Lansing, MI 48824, USA.

[c] School of Computer & Software, Nanjing University of Industry Technology, Nanjing 210016, P.R. China;

[d] Key Laboratory of Integrated Regulation and Resource Development on Shallow Lake of Ministry of Education, College of Environment, Hohai University, Nanjing 210098, P.R. China；

* Corresponding author: E-mail: zhaox26@okemosk12.net (xiabinchen Zhao),
Tel: +86 13585121654; +1 5173487286;

## Abstract

The biofilms response on active attachment surfaces (submerged plant leaves) and inert attachment surfaces (biomimetic plant glass attachment surfaces) and their effects on PAHs and nitrogen transformation under the application conditions of bactericide, algaecide and PAHs were investigated by lab simulated hydroponics and high-throughput molecular biology methods in *Vallisneria natans* (VN), *Hydrilla verticillata* (HV) and biomimetic plants (BP) systems. Results showed that the introduction of bactericide, algaecide and PAHs changed the microorganism composition in biofilm, the presence or absence of submerged plants was the primary factor causing differences in microbial communities. Moreover, the microbial diversity and endemic species of biomimetic plant biofilms were higher than those of submerged plant systems, indicating that submerged plants have selectively induced the reconstruction of biofilm-leaves. The introduction of bactericide, algaecide and PAHs leads to abnormal accumulation of TP and $NH_3$-N in overlying water, as well as $NO_3$-N and TP content in sediment. However, submerged plants can weaken and alleviate the stress effects of these factors on nitrogen and phosphorus conversion. Compared to the inert biomimetic plant glass surface, the presence of active surfaces in submerged plants results in a much higher abundance of PAHs degrading bacteria (*Sphingomonas* and *Novosphingobium*) and nitrogen converting bacteria (e.g. denitrifying bacteria of *Methylophilus, Methylotenera, Flavobacterium, Hydrogenophaga, Aquabacterium*, and nitrogen-fixing bacteria of *Rhizobium, Azoarcus, Rhizobacter, Azonexus*) in biofilm-leaves of submerged plants after the addition of bactericide, algaecide and PAHs. Coupled degradation of PAHs and nitrogen occurs, which is beneficial for the treatment of combined pollution caused by PAHs and nitrogen.

**Key Words**: Submerged plants; bactericide and algaecide; biomimetic plants; Biofilm; PAHs dissipation.

## 1. Introduction

The interaction between biofilms, plant systems, and environmental pollutants plays a crucial role in the health and sustainability of aquatic ecosystems and in improving the quality of water environment (Edwards and Kjellerup, 2013; Srivastava et al., 2017; Guo et al., 2023; Sarkar and Bhattacharjee, 2025). Biofilms, perfectly ordered collaborating microbial community immobilized in a self-synthesized matrix and associated with a single or numerous species of bacteria, algae, fungi, and archaea that form on surfaces of submerged plant, rock, and surficial sediment submerged in water, are essential in the natural degradation of pollutants and nutrient, including polycyclic aromatic hydrocarbons (PAHs), polychlorinated biphenyls (PCBs), pesticides, nitrogen and phosphorus compounds, and so on (Biswal and Malik, 2022; Gagan et al., 2023; Sarkar and Bhattacharjee, 2025). These pollutants' accumulation in water body, especially PAHs and nitrogen, causes various environmental problems including microcystin production with algae explosion, greenhouse gas release, aquatic organisms death and human health risk, are of significant concern due to their toxicity and persistence in aquatic environments (Asghar et al., 2025).

Submerged plant systems, such as aquatic macrophytes, have gained attention for their role in mitigating pollution, primarily through their interaction with biofilms and their ability to uptake and convert nutrients and contaminants (Bhatia and Goyal, 2014; Sharma and Sharma, 2022; Gagan et al., 2023). The surface properties of plant leaves, rock and sediment, including their bioactive and inert characteristics, influence biofilm formation and functionality. Bioactive surfaces, such as aquatic macrophytes and animals, which promote microbial adhesion and growth, may enhance the bioremediation processes, while inert surfaces, such as stone, sand, cement revetment and biomimetic plants, which resist biofilm attachment, may hinder the microbial response (Bryers, 1987; Drago et al., 2018; Zhao et al., 2018a; Er-rahmani et al., 2024).

The dissipation of PAHs and nitrogen compounds in submerged plant systems is a complex process that involves both microbial activity within the biofilm and the plant's physiological mechanisms. Nitrate addition could stimulate biodegradation of PAHs, which might be owing to enrichment of functional genes involved in N-, carbon (C)-, sulfur (*S*)- and phosphorus (P)- cycling processes, especially those microorganisms with diverse metabolic abilities leading to PAHs reduction/degradation(Xu et al., 2015). Meanwhile, the presence of PAHs reduced the diversity of nitrogen-fixing bacteria, and prevented the growth of many nitrogen fixing bacteria, such as proteobacteria and cyanobacteria (Sun et al., 2012).

Although recent studies presented that submerged macrophytes and their biofilms-leaves played an important role in nitrogen cycling and PAHs degradation, little is known about how different surface types—bioactive and inert—affect biofilm-leaves involved into submerged macrophytes planted in PAH-polluted sediment under the application of bactericide and algaecide. This study aims to investigate the effect of these attachment surfaces on biofilm development, the microbial response, and the dissipation rates of PAHs and nitrogen in submerged plant systems through researching the response of functional bacteria on biofilms on leaves, and combining with habitat indicators under the application of bactericide and algaecide. The aim of this study, following hypotheses had been put forward: (1) There may be a coupled degradation behavior between PAHs and nitrogen in submerged plant systems; and (2) the effect of PAHs on nitrogen cycling might be alleviated due to the application of submerged macrophytes in PAHs-polluted sediment systems, shedding light on their potential role in sustainable water treatment strategies.

## 2. Materials and methods

### *2.1 Materials*

We selected *Vallisneria natans* (VN) and *Hydrilla verticillata* (HV) (Nanjing Sam Creek aquatic breeding research base) as the bioactive surface and Bionic plant (BP) with similar glass surface area as the inert surface. Sediments (pH=7.38, organic matter=2.08%, the background value of phenanthrene (Phe) and Pyrene (Pyr) level are 0.028 mg/kg and 0.013 mg/kg, respectively) were collected from a suburb river of Nanjing (not in the main industrial area), air-dried, manually crushed, and then sieved with 2-mm mesh to remove plant residues and stone. Glass containers (39 diameter × 51 cm height) were chosen to cultivate submerged plants, to avoid loss of Phe and Pyr via adsorption. The experiment was carried out in the ecological greenhouse with three replicates for 35 d.

### *2.2 Experimental setup*

The 0.84 g Phe and 0.056 g Pyr dissolved in acetone (1000 mL) was spiked into 3.5 kg sediment. After acetone evaporating, the polluted sediment was mixed with unpolluted sediment with their respective proportion, and laid in each container smoothly. The final contents of Phe and Pyr in sediment (dry weight) were 30 mg/kg and 2 mg/kg, respectively.

The sterilizing agent (about 0.01 g norfloxacin and 0.015 g roxithromycin / L water) were added into water to remove or destroy the leaf-surface biofilm. After the completion of plant domestication (3-5 d), healthy plants were transplanted into the containers. 50 L water was added to the container and the water-line was marked clearly to replenish water to a uniform level throughout cultivation.

Thirteen glass buckets with a diameter of 39 cm and a height of 51 cm were selected for the experiment to cultivate submerged plants, which were divided into four major systems, (1) *Vallisneria natans* (VN): VN, VN+PAHs, VN+PAHs+bactericide, VN +PAHs+algaecide, and VN+PAHs+bactericide and algaecide; (2) *Hydrilla verticillata* (HV): HV, HV+PAHs, HV+PAHs+bactericide, HV+PAHs+algaecide, and HV+PAHs+bactericide and algaecide; (3) Biomimetic plant (BP): BP+PAHs, and BP+PAHs+ bactericide and algaecide; (4) Control group: also known as sediment control group. The concentration of bactericide and algaecide were 1 mg norfloxacin / L water + 15 mg roxithromycin / L water and 2.5 mg $CuSO_4$ / L water, respectively.

Water and sediment samples were collected for Phe and Pyr analysis at 14, 28 and 36 d. The samples were stored at -20°C for PAHs analysis. The leaf-surface biofilms of submerged plants and Biomimetic plants were extracted in 14 and 28 d, respectively.

### *2.3 The separation of biofilm attached on the plants leaves*

The method of biofilm separation from leaf surfaces was modified from Zhao et al.(2018a). With precool ethanol-PBS buffer as eluent, appropriate amount of leaves were put in the polyethylene test tube, then Triton solution and several 3 mm glass beads were added. All the sample tubes were placed in reciprocating oscillator with constant temperature and were shaken for 10 min (225 r/min), then underwent ultrasounds (150 W, 40 kHz) for 1 min. After filtrating elution liquor, the filtrate was centrifuged for 10 min (10000 rpm), and the centrifugal precipitate was collected. Biomimetic plant biofilms could be scraped directly with a sterile scalpel.

### *2.4 Analysis methods*

*2.4.1 Extracting, purifying and determination of PAHs from sediment and water samples*

The extracting and purifying of PAHs from sediment are ultrasonic extraction method and SPE column purifying method (Zhao et al., 2018a). The PAHs in water extracted with C18 SPE column method (Zhao et al., 2018b).The obtained eluent was concentrated and adjusted to 1 mL with *n*-hexane for HPLC determination. The phe and Pyr were analyzed by Agilent1100 HPLC with fluorescence and UV - adsorption detector. Their concentrations were quantified by using external standard solutions sourced from Ehrenstorfer (Augsburg, Germany). Detective wavelength with FLD signals: Exλ=257 nm, Emλ=380 nm (Zhao et al., 2018b).

*2.4.2 DNA extraction,*

Dissolved oxygen (DO), pH, oxidation-reduction potential (ORP) and conductivity (EC) were measured in situ by using SX 751 series portable electrochemical meters (SanXin Instrumentation Inc., Shanghai, China). Total nitrogen, nitrate nitrogen and ammonia nitrogen in water (TN, $NO_3^-$-N, $NH_4^+$-N) were measured in the laboratory with 24 h after fixing water samples with concentrated sulfuric acid; TN, $NO_3^-$-N and $NH_4^+$-N were measured by potassium persulfate oxidation-ultraviolet spectrophotometry method, ultraviolet spectrophotometric screening method and Nessler's reagent spectrophotometry, respectively. Sediment was frozen and dried for one week and sieved by 2-mm mesh. Total nitrogen, nitrate nitrogen and ammonia nitrogen of sediment (TN, $NO_3^-$-N, $NH_4^+$-N) were analyzed by hypobromate oxidation method, ultraviolet spectrophotometry and indiphen blue colorimetric method, respectively. TN of plants was measured by micro kjeldahl method (Zhao et al., 2018a).

*2.4.4 DNA extraction, PCR amplification, sequencing and data analysis*

Weighing 0.5 g samples in 2 mL centrifuge tube, the bacterial DNA on the leaves surface were extracted by Soil DNA Kit (Omega E.Z.N.A.$^{TM}$, Omega Bio-Tech) according to manufacturer's protocol. The PCR primers were V3-V4 universal primers 341F/805R (341F: CCTACGGGNGGCWGCAG; 805R: GACTACHVGGGTATCTAATCC) provided by Sangon Biotech Co., Ltd., Shanghai, China. The PCR reaction mixture contained 5 μL 10×PCR buffer, 0.5 μL dNTPs (10 mM each), 0.5 μL Bar-PCR primer F (50 μM), 0.5 μL Primer R (50 μM), 0.5 μL Plantium Taq (5 U/μL), 10 ng DNA template, with sterile water added to make the final volume to 50 μL. The following PCR cycle was performed: initial denaturation at 94 °C for 3 min, denaturation at 94 °C for 30 s, renaturation at 94 °C for 30 s, annealing at 45°C for 20 s, extension at 65 °C for 30 s with a total of 30 cycles, and the final extension at 72 °C for 5 min. Amplification products were detected by 1% agarose gel electrophoresis, and then recycled using DNA Recycle Kit, SK8131 (Sangon Biotech Co., Ltd, Shanghai, China), and finally quantified using Qubit2.0 DNA Assay Kit (Sangon Biotech Co., Ltd, Shanghai, China). Paired-end sequencing was performed using Illumina MiSeq platform (Illumina, San Diego, CA, USA).

Sequencing analysis was executed by using QIIME (Yu et al., 2017). The Uclust method was used to pick the representative sequences for each operational taxonomic units (OTUs). Then the taxonomic information was annotated with the SILVA database after subsampling based on the lowest number of reads. Statistical analyses were performed using R software, including rank-abundance curves, microbial community composition at phylum and genus level, NMDS analysis using Bray-Curtis algorithm. Heat maps were generated by HemI. The difference of microbial communities at genus level was analyzed by Kruskal-wallis H test and Wilcoxon rank-sum test using STAMP software.

*2.5. Statistical analysis*

The correlation analysis of ORP, DO, pH, COND, water quality index and the residual concentration of Phe and Pyr were analyzed by two-way ANOVA in SPSS 23 software for windows at a significant level of $p < 0.05$.

In this study, the experimental data were analyzed using Redundancy Analysis (RDA). The analysis was implemented in Python utilizing the rpy2 library to interface with the vegan package in R. This approach allows for the simultaneous visualization of relationships among samples, environmental factors, and microorganisms, as well as pairwise correlations between any two of these components. The method enables the identification of significant environmental factors influencing sample distribution and helps detect key species associated with environmental variations. The pseudocode for this analytical approach is as follows:

Pseudocode for RDA Analysis：

```
#Data Preprocessing: Hellinger Transformation of Species Data
FUNCTION hellinger_transform(df):
    CALCULATE row sums
    DIVIDE each row by its row sum
    TAKE square root of the result
    RETURN transformed data
sp_hellinger = hellinger_transform(sp_data)
# Check for Multicollinearity Among Environmental Variables. If environmental factors are #highly correlated, results may be distorted. Variance Inflation Factor (VIF) can be used to remove #factors with VIF > 10
FUNCTION calculate_vif(env_df):
    CONVERT env_df to R object
    IN R:
        CREATE model matrix
        CALCULATE VIF values using car::vif
    RETURN dictionary of VIF values
vif_values = calculate_vif(env_data)
# Perform RDA Analysis
FUNCTION perform_rda(species_df, env_df):
    CONVERT species_df and env_df to R objects
    IN R:
        EXECUTE vegan::rda(species_data ~ env_variable1 + env_variable2 + ..., data=environment_data)
    RETURN RDA result object
rda_result = perform_rda(sp_hellinger, env_data)
# Extract Results and Calculate Variance Explanations
EXTRACT eigenvalues from rda_result
CALCULATE total_variance = sum of all eigenvalues
CALCULATE constrained_variance = sum of constrained axis eigenvalues
```

CALCULATE explained_variance_ratio = constrained_variance / total_variance
# Statistical Testing and Visualization
TEST statistical significance
EXTRACT scores for visualization
VISUALIZE results (biplot and scree plot).

## 3 Results and Discussion

***3.1 Characteristics of changes in physical and chemical indicators of water bodies***

Fig.1 showed the variation characteristics of ORP, DO, pH, EC in different aquatic systems, as well as TN, ammonia nitrogen, and TP in overlying water. According to Fig. 1, the variation range of ORP under different treatment conditions during the experiment was 20 -140 $mV$, with higher ORP values in the biomimetic system and control group than in the submerged plant system. During the experiment, there were significant fluctuations in DO values under different treatment conditions, ranging from 0.39 to 8.88 mg/L. This may be related to plant oxygen release, decay and regeneration, as well as microbial growth and extinction during the experiment (Qin et al., 2019). The pH values of submerged plant system were higher than those of the blank and biomimetic plant control groups, and this pattern was particularly evident in the HV system. This may be related to the consumption of $CO_2$ in water by submerged plants through photosynthesis (Liu et al., 2016). The EC values of the biomimetic system and the control group were higher than those of the submerged plant system. It can be seen that the presence of submerged plants and microorganisms can consume or assimilate a large amount of ionic compounds, and have an impact on the physical and chemical environment of water bodies.

The presence of PAHs in biomimetic systems has caused a certain degree of blockage and accumulation of TP and ammonia nitrogen conversion in the overlying water; However, the effect of PAHs in the plant system is significantly weakened, which may be related to the absorption and transformation of $NH_3$-N by submerged plants and microorganisms, weakening and alleviating the stress effect of PAHs (Qin et al., 2006; Chen et al., 2013). The inhibitory effects of PAHs on the transformation of $NH_3$-N and TP in overlying water showed differences in HV system and VN system, which may be related to the differences in growth rate, nutrient demand and absorption capacity, and promotion of microbial growth between the two submerged plants (Fraser et al., 2004; Lee et al., 2010). Similarly, compared to plant system, $NO_3$-N and TP accumulated significantly in sediment of non-plant system (BP), while $NH_3$-N and TN did not accumulate significantly in sediment. The presence or absence of submerged plants has little effect on the content of Phe in water, but has a significant impact on the content of Pyr. The VN system significantly reduces the content of Pyr in overlying water. Interestingly, compared to biomimetic plant systems (BP), the presence of submerged plants to some extent promotes the degradation of Phe and Pyr in sediments (Fig.1), which may be related to the well-developed roots and root exudates of submerged plants (Qin et al., 2022). In addition, we also found that the degradation rate of Phe in sediments (about 70.86%~88.61%) was higher than that of Pyr (about 61.22%~82.85%), which is consistent with previous research results (Lee et al., 2008; Meng and Chi, 2015).

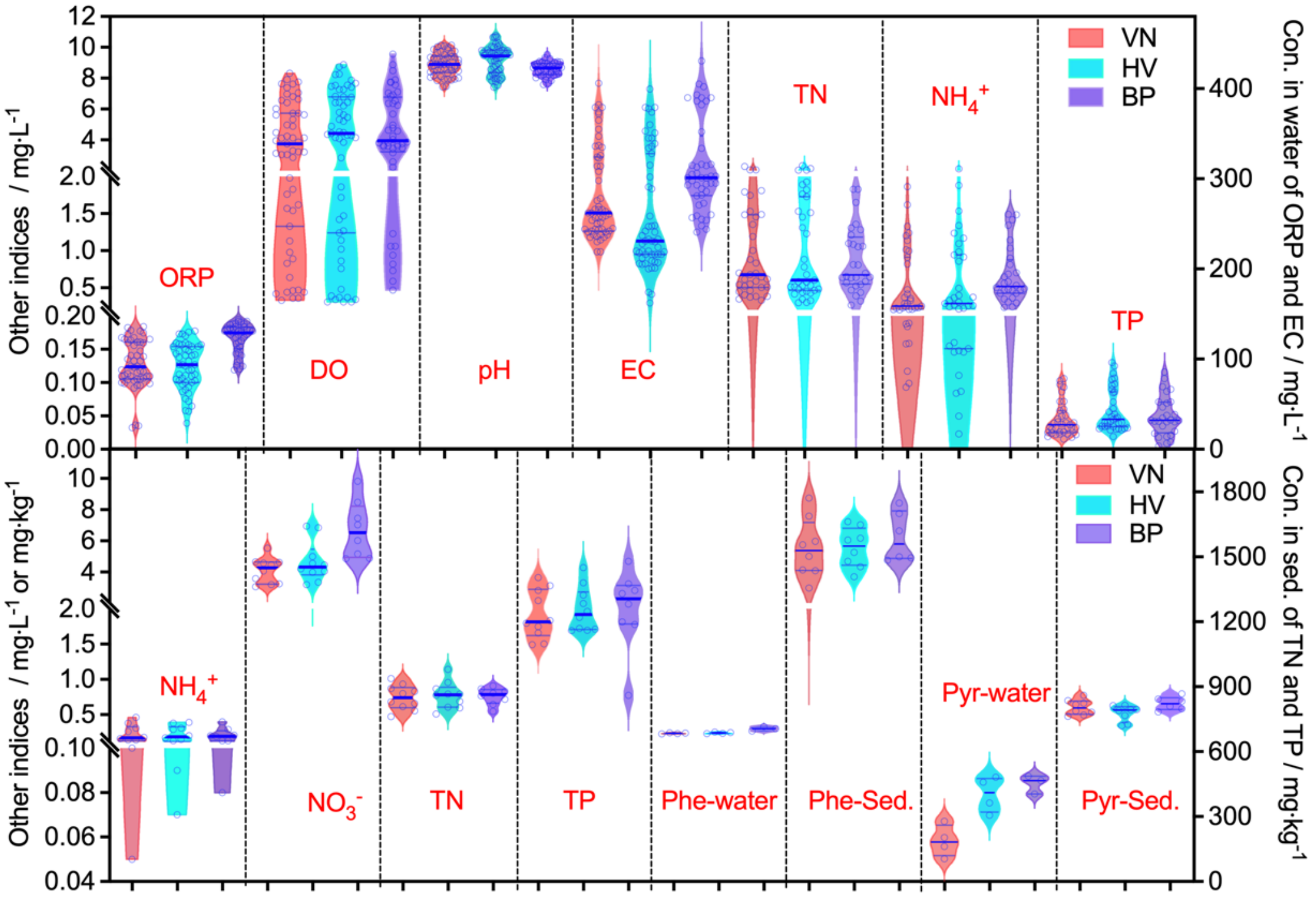

**Fig.1** Characteristics of changes in physical and chemical properties and water quality indicators in water bodies and sediment. Note: (1) VN: *Vallisneria natans (Lour.) Hara*, HV: *Hydrilla verticillata*, BP: Bionic plant. (2) Phe / Pyr-water: concentration of phenanthrene / pyrene in water. Phe / Pyr-Sed: concentration of phenanthrene / pyrene in sediment.

***3.2 Correlation analysis between the residual concentration of Phe and Pyr and environmental factors***

Pearson and Spearman correlation analysis were conducted on the physicochemical indicators (ORP, DO, pH, and EC), water quality indicators (TN, $NH_3$-N, and TP in water), TN, $NH_3$-N, $NO_3$-N, TP in sediments, and the residual concentrations of Phe and Pyr in sediments, to explore the main factors affecting the degradation rates of Phe and Pyr. According to Fig.2, the residual concentrations of Phe and Pyr in sediment are positively correlated with ORP, EC, $NH_3$-N in water and $NH_3$-N, $NO_3$-N in sediment, and negatively correlated with DO in water. Among them, the concentration of Pyr in water was significantly positively correlated with ORP in water ($P<0.05$); The concentration of Phe in water is significantly negatively correlated with DO in water ($P<0.05$) and significantly positively correlated with EC ($P<0.05$); The concentrations of Phe and Pyr in water were significantly positively correlated with $NH_3$-N in sediment ($P<0.05$); In addition, there was a highly significant positive correlation ($P<0.01$) between the residual concentrations of Phe and Pyr in sediment and water. This indicates that DO in water plays an important role in the degradation of polycyclic aromatic hydrocarbons (PAHs), possibly because the microorganisms involved in PAHs degradation require aerobic or microaerophilic environments (Martiranivon et al., 2017). The nitrogen content in water and sediment plays an important role in the reduction of PAHs in sediment, and there may be a coupled degradation of nitrogen and PAHs (Macrae and Hall, 1998; Wang et al., 2012; Yang et al., 2013). It is worth noting that there is a highly significant positive correlation between the residues and degradation of Phe and Pyr in water and sediment, indicating that the residues and degradation of Phe and Pyr are complementary and mutually influential processes, and may involve co-metabolism (Dan et al., 1999).

In addition, Fig.2 also reflects the good correlation between environmental factors. DO in water is significantly positively correlated with ORP and pH ($P$<0.05), and extremely significantly negatively correlated with EC ($P$<0.01), while $NO_3$-N in sediment is significantly negatively correlated with pH and DO ($P$<0.01).

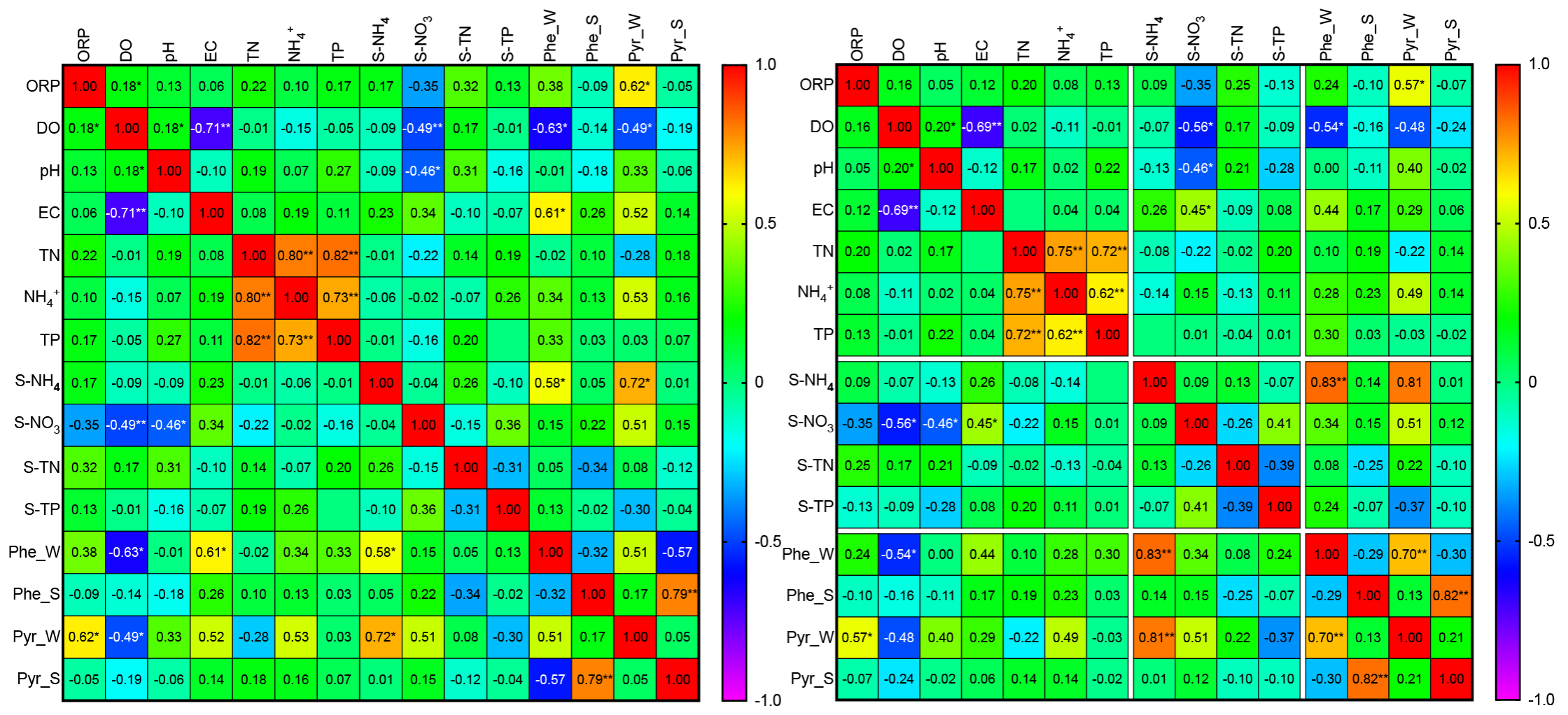

**Fig.2** Correlation heatmap of physical and chemical factors based on Pearson (left) and Spearman (right) correlation. Note: (1) S-$NH_4$, S-$NO_3$, S-TP: the concentration of $NH_4^+$, $NO_3^-$ and TP in sediment; Phe_W, Pyr_W, Phe_S and Pyr_S: the concentration of Phe and Pyr in water and sediment. (2) The "*" and "**" represent significant under $P < 0.05$ and $P < 0.01$level, respectively.

### *3.3 Microbial response of biofilm-leaves*

In order to better study the changes in microbial diversity in water systems treated with five types of bactericide and algaecide methods, we used high-throughput methods to compare the structural composition and diversity changes of microorganisms in foliar biofilms in each system.

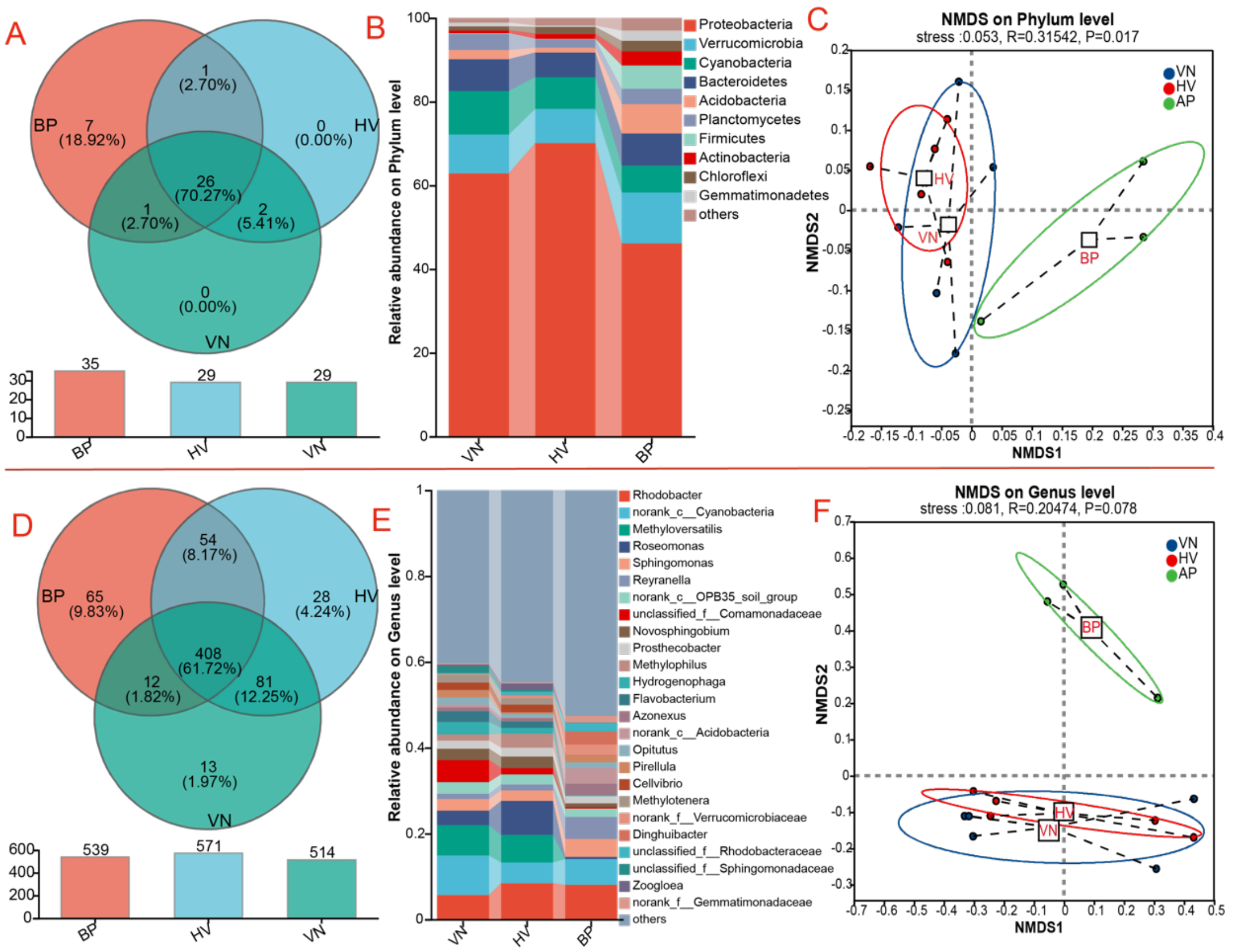

**Fig.3** The Venn graph (Fig. A and B), column graph (Fig. B and E) and NMDS graph (Fig. C and F) of biofilm-leaf composition on the surface of BP, HV and VN under phylum level (Fig. A, B and C) and genus level (Fig. D, E and F). Note: (1) VN: *Vallisneria natans (Lour.) Hara*, HV: *Hydrilla verticillata,* BP: Bionic plant.

The Fig.3 showed the Venn graph, column graph and NMDS graph of biofilm-leaves composition on the surface of BP, HV and VN at phylum level and genus level. From Fig.3C and Fig.3F, it can be seen that there are significant differences in the microbial composition of biofilms in HV, VN, and BP, especially in plant biofilm-leaves and BP biofilm. It can be found that plant active leaves have a strong induction effect on the microbial composition of their surface biofilms, while plant species have little effect on microbial composition. Fig.3A and Fig.4D show that BP has the highest number of microbial phyla (35 phyla), while the induction effect of plant active leaves leads to a significant decrease in microbial phyla in the biofilm-leaves of HV and VN (29 phyla). In terms of genus level, the induction effect of active leaves is not significant, and there is no significant difference in the number of genus species between BP biofilm and biofilm-leaves (514-571). However, the number of unique species in BP biofilm (65) is significantly higher than that in biofilm- leaves (13-28). Fig.3B and Fig.3E indicate that the main differences between plant bioactive leaf surfaces and biomimetic plant inert surfaces at the phylum level are mainly due to Proteobacteria. In terms of genus level, there are significant differences in the composition of microbial communities among different treatment samples.

Among them, *unclassified_f_Ruminococcaceae*, *Haloferula,* *Dinghuibacter*, *norank_f_*

*Gemmatimonadaceae*、*Pseudoxanthomonas*）are the unique bacterial community in the biofilm of biomimetic plant system and are not abundant. While *Rhodobacter*, *norank_c_ Acidobacteria*, *Reyranella*, *norank_f_Verrucomicrobiaceae*, *Azonexus*, *Sphingomona and unclassified-f-Rhodobacteraceae*It are its advantageous microbial community. For submerged plant biofilm-leaves, *Methyloversatilis, Roseomonas, Comamonadaceae, Methylophilus, Cellvibrio* 、*Methylotenera, Hydrogenophaga, Flavobacterium, Sphingomonadaceae, Meganema, Roseococcus, Acidibacter, Methylophilaceae, Zoogloea are their unique and advantageous bacterial genus.* This may be related to the fact that submerged plants can secrete corresponding substances to induce microbial composition (Cufrey and Kemp, 1992; Hernandez and Mitsch, 2015; Qin et al., 2022)。Comparing the composition of microorganisms attached to the leaves of VN and HV systems, it can be found that, in addition to the abundance differences between the genera of *norank_c_Cyanobacteria* (HV： VN=7:13), *Roseomonas* (HV: VN=11:5) and *Rhodobacter* (HV: VN=3:2) are significant, the abundance of their bacterial communities is generally similar.

In order to evaluate the significance level of species abundance differences, and obtain species with significant differences between samples, strict statistical methods are used to conduct hypothesis tests on species between different groups of microbial communities based on the obtained community abundance data. The top 50 species in abundance among multiple samples, Kruskal Wallis rank sum test was selected. The results showed that for different attachment surfaces, there was a highly significant difference in the abundance of *norank_f_ Acidobacteria* among the samples ($P$<0.01), while the abundance of *unclassified d_f_Sphingomonadaceae*, *Methylloversatilis*, *Methylobacteria*, *Hydrophages*, *Vibrio fibrosus*, *Methylotener, norank_f_ Gemmatimondaceae* and other species showed significant differences among the samples ($P$<0.05).

### *3.4 Identification of PAHs and nitrogen converting bacteria in foliar biofilms-leaves and their relationship with environmental factors*

Fig.4A and Fig.4B depicted the relationship between the residual concentration of Phe and Pyr, environmental factors and bacteria at phylum level (Fig.4 A) or genus level (Fig.4 B) in biofilm-leaves with Redundancy Analysis (RDA). At the phylum level, redundancy analysis was conducted between the top 10 phyla bacterial communities in terms of their abundance, environmental foctors and the residual concentration of Phe and Pyr. Results showed that. Proteobacteria, Cyanobacteria, Cyanobacteria, and Verrucomicobia have the greatest impact on the occurrence of Phe / Pyr and nitrogen / phosphorus in the environment, followed by Bacteroidetes, Planctomycetes, Firmicutes, and Acidobacteria, while the impact of other bacterial phyla is relatively small. Proteobacteria, Verrucomibia, and Bacteroidetes have the greatest impact on the occurrence of Phe and Pyr in water and sediment. As for the contribution of microorganisms at the genus level, *Methylloversatilis, Rhodobacter, Sphingomonas,* and *Roseomomonas* have the greatest impact on the occurrence of Phe / Pyr and nitrogen / phosphorus in the environment, followed by *Novosphingobium, Flavobacterium, Prosthecobacter, Reyranella, norank_c__Cyanobacteria, unclassified_f__Comamonadaceae,* and *Hydrogenophaga*. Among them, *Methyloversatilis, Rhodobacte, Sphingomonas, Roseomonas,*

*Novosphingobium, Flavobacterium, Prosthecobacter, Reyranella* have the greatest impact on the occurrence of Phe and Pyr in water and sediment.

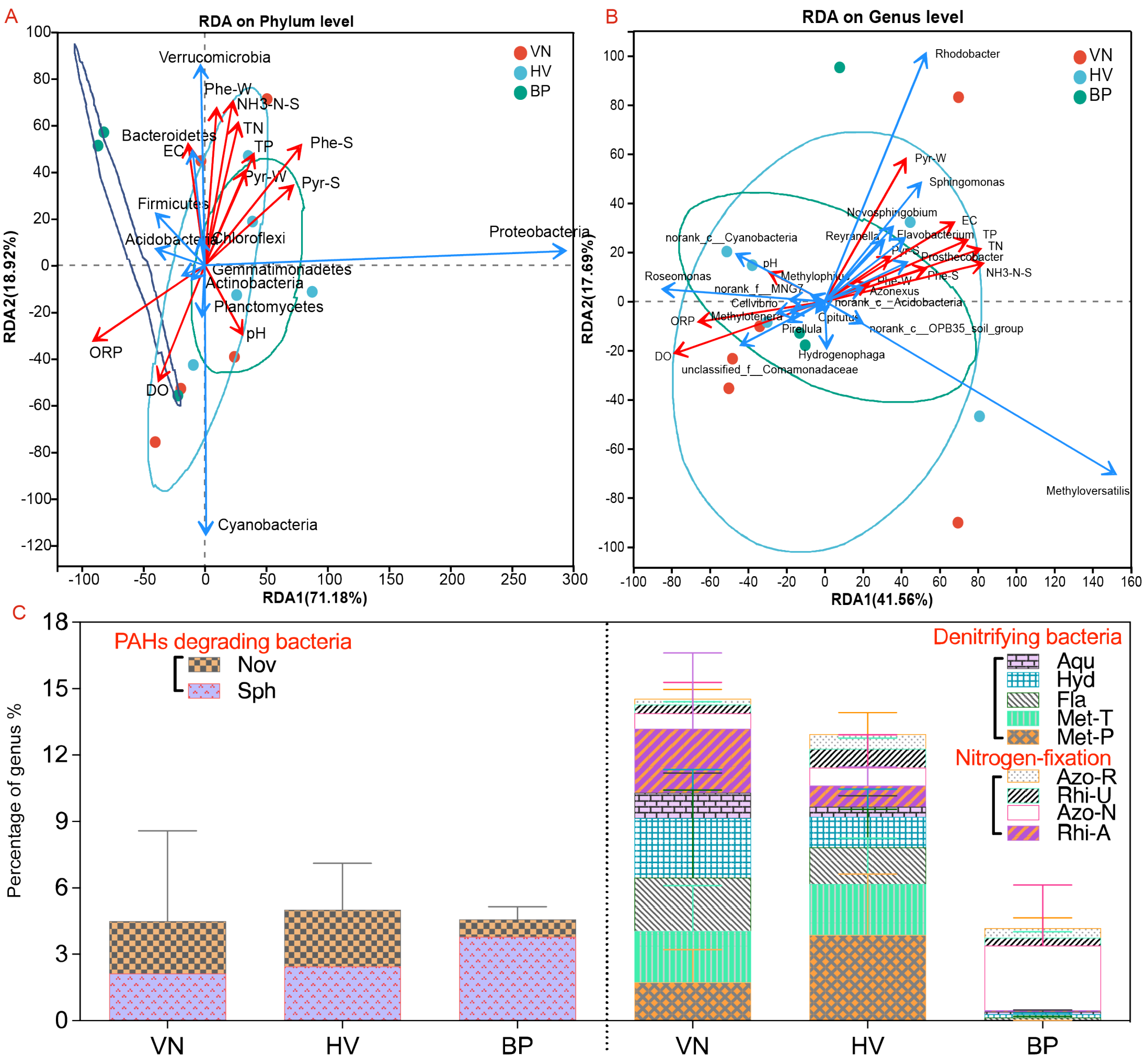

**Fig.4** The Redundancy Analysis (RDA) depicted the relationship between the residual concentration of phenanthrene and pyrene, environmental factors and bacteria at phylum level (Fig.5 A) or genus level (Fig.5 B) in biofilm-leaves. The Fig.5C showed the percentage of genus with PAHs degrading and nitrogen transformation function. The circles represent the collected biofilm samples with different attachment surfaces. The blue arrows mean bacterial taxa at genus or phylum level and the red arrows mean the environmental factors.Note: (1) VN: *Vallisneria natans (Lour.) Hara*, HV: *Hydrilla verticillata,* BP: Bionic plant. (2) $NH_3$-N-S: the concentration of $NH_4^+$ in sediment; TN, TP: the concentration of $TN^-$ and TP in water; Phe-W, Pyr-W, Phe-S and Pyr-S: the concentration of Phe and Pyr in water and sediment. (3) Sph: *Sphingomonas,* Nov: *Novosphingobium,* Met-P: *Methylophilus,* Met-T: *Methylotenera,* Fla: *Flavobacterium,* Hyd: *Hydrogenophaga,* Aqu: *Aquabacterium,* Rhi-U: *Rhizobium,* Azo-R: *Azoarcus,* Rhi-A: *Rhizobacter,* Azo-N: *Azonexus.*

We screened and identify those that can degrade PAHs and participate in nitrogen conversion bacteria from the top 50 bacterial communities with abundance at the genus level. Fig.4C showed the diversity of PAH degrading bacteria and nitrogen converting bacteria in the top 50 bacterial communities with abundance at the genus level. Observing the composition and abundance of functional bacteria in the samples, it can be found that the different sterilization treatments and attachment surfaces result in a grouping phenomenon, and the abundance of functional bacteria on the active attachment surface (submerged plants) is significantly higher than that on the inert attachment surface (biomimetic plants glass). This may be because submerged plants secrete substances that are beneficial for microbial growth (Cufrey and Kemp, 1992; Hernandez and Mitsch, 2015).

In biomimetic plant systems, the addition of bactericide, algaecide and PAHs disturbed the microbial community to a certain extent, and the abundance of most nitrogen converting microorganisms (such as *Azonexus* and *Azoarcus*) decreased by three times or even hundreds of times. In the submerged plant system, for VN and HV, the addition of bactericide, algaecide and PAHs increased the abundance of nitrogen-fixing dominant bacteria and denitrifying bacteria, which were much higher than those in the biomimetic plant system. This indicates that the presence of submerged plants provides a good living environment for these nitrogen converting microorganisms, such as denitrifying bacteria of *Methylophilus, Methylotenera, Flavobacterium, Hydrogenophaga, Aquabacterium*, and nitrogen-fixing bacteria of *Rhizobium, Azoarcus, Rhizobacter, Azonexus*.

For the microbial community capable of degrading PAHs, in the biomimetic plant system, the addition of PAHs increased the abundance of PAH degrading bacteria (Lyu et al., 2014) and *Sphingomonas* (Cébron et al., 2008) by 2-3 times. Indicating that the presence of PAHs induces the growth and reproduction of this microbial community. After adding bactericide and algaecide simultaneously, the abundance of PAH degrading bacterial communities increased by 5-10 times. This may be due to the inhibitory effect of bactericide and algaecide on other bacterial communities, it provides an advantage for the growth and reproduction of PAH transforming bacteria, and occupies favorable conditions in competition. In the submerged plant systems, the abundance of *Sphingomonas* decreased after the addition of PAHs, while the abundance of *Novosphingobium* increased. However, after the addition of bactericide and algaecide, the abundance of both bacterial groups significantly increased, which is consistent with the pattern in biomimetic plant systems.

## 4. Conclusion

This article mainly uses lab simulated hydroponics and high-throughput molecular biology methods, using submerged plants such as VN, HV, and biomimetic plants (BP) as models, to study the response of biofilms on active attachment surfaces (submerged plant leaves) and inert attachment surfaces (biomimetic plant glass attachment surfaces) under the conditions of bactericide, algaecide and PAHs, and their effects on PAHs and nitrogen transformation in sediment and water. Results showed that the introduction of bactericide, algaecide and PAHs would have an impact on the microorganism composition in biofilm, but the presence or absence of submerged plants was the primary factor causing differences in microbial communities. Moreover, the microbial diversity and endemic species of biomimetic plant biofilms were higher than those of submerged plant systems, indicating that submerged plants have selectively induced the reconstruction of biofilm-leaves. The introduction of bactericide, algaecide and PAHs leads to abnormal accumulation of TP and ammonia nitrogen in overlying water, as well as $NO_3$-N and TP content in sediment. However, submerged plants can weaken and alleviate the stress effects of these factors on nitrogen and phosphorus conversion. Compared to the inert biomimetic plant glass surface, the presence of active surfaces in submerged plants results in a much higher abundance of PAH degrading bacteria and nitrogen converting bacteria in biofilm-leaves of submerged plants after the addition of bactericide, algaecide and PAHs. Coupled degradation of PAHs and nitrogen occurs, which is beneficial for the treatment of combined pollution caused by PAHs and nitrogen.

**Acknowledgments:** This research was funded by the Industry's Talent Introduction Research Initiation Fund Project (Natural Science Category) of Nanjing Vocational and Technical University (No. YK22-05-02) and Major Natural Science Research Projects of Jiangsu Province Universities (No. 23KJA520008), National Natural Science Foundation of China (Grants No. 51509129, 51879080 and 41371307).